\documentclass[superscriptaddress,amsmath,amssymb,nofootinbib,eqsecnum,a4paper,secnumarabic,11pt]{revtex4}    
\usepackage{graphicx}
\usepackage{hyperref}  
\usepackage{titlesec}    
\usepackage[english]{babel}

\begin{document}
\title{Classification of simple heavy vector triplet models}
\author{Tomohiro Abe}
\affiliation{
Institute for Advanced Research, Nagoya University,
Furo-cho Chikusa-ku, Nagoya, Aichi, 464-8602 Japan
}
\affiliation{
Kobayashi-Maskawa Institute for the Origin of Particles and the
Universe, Nagoya University,
Furo-cho Chikusa-ku, Nagoya, Aichi, 464-8602 Japan
}
\author{Ryo Nagai}
\affiliation{
Department of Physics,  Nagoya University,
Furo-cho Chikusa-ku, Nagoya, Aichi, 464-8602 Japan
}

\begin{abstract}
We investigate decay modes of spin-1 heavy vector bosons ($V'$) from the viewpoint of perturbative unitarity
in a model-independent manner.
Perturbative unitarity requires some relations among couplings.
The relations are called unitarity sum rules.
We derive the unitarity sum rules from processes that contain two fermions and two gauge bosons.
We find the relations between $V'$ couplings to the SM fermions $(f)$ and $V'$ couplings to the SM gauge bosons ($V$).
Using the coupling relations, we calculate partial decay widths for $V'$ decays into $VV$ and $ff$.
We show that Br($W' \to WZ) \lesssim$ 2\% in the system that contains $V'$ and CP-even scalars as well as the SM particles.
This result is independent of the number of the CP-even scalars. 
We also show that contributions of CP-odd scalars help to make Br($W' \to WZ$) larger than Br($W' \to ff$)
as long as the CP-odd scalars couple to both the SM fermions and the SM gauge bosons.
The existence of the CP-odd scalar couplings is a useful guideline to construct models that predict Br($W' \to WZ) \gtrsim$ 2\%.
Our analysis relies only on the perturbative unitarity of $f\bar{f} \to WW'$.
Therefore our result can be applied to various models.

\end{abstract}

\maketitle

\section{Introduction}

The ATLAS and the CMS experiments at the Large Hadron Collider (LHC)
are capable of discovering physics beyond the standard model (BSM) around the TeV scale. 
In 2012, the LHC successfully discovered a scalar particle~\cite{Aad:2012tfa, Chatrchyan:2012xdj} 
that is consistent with the Higgs boson predicted in the standard model (SM). 
The next target of the LHC is particles predicted in BSM models.

Spin-1 heavy vector particles ($V'$) are popular particles predicted in BSM models such as
composite Higgs models~\cite{0709.0007,0810.1497,0911.0059,1109.1570,1205.4032,1208.0268}.
An efficient way to study $V'$ phenomenology at the LHC is to use effective Lagrangians.
For example, the effective Lagrangian given in Refs.~\cite{0907.5413,1402.4431}
provides a simple framework for $V'$ phenomenology at the LHC,
and the framework is used in the analysis of ATLAS and CMS~\cite{1406.4456, 1502.04994, 1503.08089, 1506.01443, 1601.06431, 1606.04833}.
On the other hand, the effective Lagrangian violates perturbative unitarity at higher energy scales in general.
If the unitarity violation scale is as low as the TeV scale, we have to take into account a lot of higher-dimensional operators. 
The higher dimensional operators accompany unknown coefficients, which make it difficult to give a definite prediction.

We can avoid the perturbative unitarity violation if the Lagrangian is renormalizable. 
The effective Lagrangian given in Refs.~\cite{0907.5413,1402.4431} includes a renormalizable model called HVT model A
in special regions of parameter space.
This model predicts that $V'$ particles mainly decay into two SM fermions ($f$)
and the decay mode into two SM gauge bosons ($V$) is much smaller: Br$(V' \to VV) \ll$ Br$(V' \to ff)$.
However, we do not know the main decay mode of $V'$ in advance of the discovery.
$V'$ could be discovered in the $VV$ decay mode in the future.
It is thus important to prepare for discovery in any decay mode, 
and thus we should prepare other Lagrangians that predict Br$(V' \to V V ) \geq$ Br$(V' \to ff )$ without violating perturbative unitarity.

In this paper, we investigate decay modes of spin-1 heavy vector bosons 
from the viewpoint of perturbative unitarity in a model-independent manner.
Our purpose is to figure out the conditions under which Br($V' \to VV) \geq $ Br($V' \to ff$) 
without relying on specific models. 
To this purpose, we need to know the $V'$ couplings to the gauge bosons and to the fermions.
We can find coupling relations
by imposing perturbative unitarity on scattering amplitudes that contain both the gauge bosons and the fermions.
These relations are called unitarity sum rules. 
Using the unitarity sum rules, we can obtain the coupling relations to calculate Br($V' \to VV)$ and Br($V' \to ff$). 
The unitarity sum rules depend on the matter contents of the system,
and thus we can understand what kinds of matter contents and interactions can make 
Br($V' \to VV)$ larger than Br($V' \to ff$) from the unitarity sum rules.

The rest of this paper is organized as follows:
In Sec.~\ref{sec:PU}, we derive the unitarity sum rules for the process where two fermions are in the initial state and
two charged gauge bosons are in the final state. The amplitudes contain terms that are proportional to energy and energy squared
in the high-energy limit. We require that these terms vanish, and obtain the unitarity sum rules.
In Sec.~\ref{sec:case1}, we analyze a system that contains $V'$ and an arbitrary number of CP-even scalars as well as the SM particles.
This system includes HVT model A. 
Using the sum rules, we investigate the ratio of the two branching ratios, Br($W' \to WZ$) and Br($W' \to ff$).
We show that the system predicts Br($W' \to WZ) \lesssim$ 2\% without depending on the number of CP-even scalars. 
We need other particles to make Br($W' \to WZ)$ larger than Br($W' \to ff$) without violating perturbative unitarity.
In Sec.~\ref{sec:case2}, we add CP-odd scalar bosons to the system, 
and show that the contributions of the CP-odd scalars significantly modify the relation between the two branching ratios.
We summarize our discussion in Sec.~\ref{sec:summary}.

\section{Unitarity sum rules for $f_1 \bar{f}_2 \to V^{-}_3 V^{+}_4$}
\label{sec:PU}

In this section, we calculate the scattering amplitude of $f_1 \bar{f}_2 \to V_3^{-} V_4^{+}$
denoted by $i {\cal M}_{jk}$, where $j$ and $k$ are twice the helicity of $f_1$ and $\bar{f}_2$, respectively.
We include 
all the SM fermions, 
the SM gauge bosons ($W^{\pm}, Z, \gamma$),
new heavy vector bosons ($W'^{\pm}, Z'$),
CP-even scalars ($h$'s),
and CP-odd scalars ($\Delta^{0}$'s)
in our analysis.
We assume CP conservation in the scalar sector.

We begin by listing all the relevant interaction terms.
The quark couplings to the gauge bosons are
\begin{align}
 {\cal L}_{ffV}
=& 
- \sum_{V = W, W'} \bar{u}_i \gamma^{\mu} V^+_{\mu} 
\left( \frac{1}{\sqrt{2}} g_{\bar{u}_i d_j V^+}^L P_L  \right)
d_j
- \sum_{V = W, W'}  \bar{d}_j \gamma^{\mu} V^-_{\mu} 
\left( \frac{1}{\sqrt{2}} (g_{\bar{u}_i d_j V^+}^L)^{*} P_L \right)  u_i
\nonumber\\
&
- \sum_{V = Z, Z', \gamma} 
  \bar{u}_i \gamma^{\mu} V^0_{\mu} (g_{\bar{u}_i u_j V^0}^L  P_L + g_{\bar{u}_i u_j V^0}^{R}  P_R ) u_j
\nonumber\\
&
- \sum_{V = Z, Z', \gamma} 
  \bar{d}_i \gamma^{\mu} V^0_{\mu} (g_{\bar{d}_i d_j V^0}^L  P_L + g_{\bar{d}_i d_j V^0}^{R}  P_R ) d_j
,
\label{eq:ffV}
\end{align}
where 
$g_{\bar{u}_i u_j V^0}^{L,R} = (g_{\bar{u}_j u_i V^0}^{L,R})^{*}$, 
and $g_{\bar{d}_i d_j V^0}^{L,R} = (g_{\bar{d}_j d_i V^0}^{L,R})^{*}$ to keep Hermiticity of the Lagrangian.
The quark couplings to the neutral scalar bosons are \begin{align}
 {\cal L}_{ffS}
=&
- \sum_h h  \bar{u}_i (g_{\bar{u}_i u_j h} ) u_j  
- \sum_h h  \bar{d}_i (g_{\bar{d}_i d_j h} ) d_j  
\nonumber \\ 
& 
+ i \sum_{\Delta^0} \Delta^0  \bar{u}_j (g_{\bar{u}_j u_k \Delta^0} \gamma^5) u_k
+ i \sum_{\Delta^0} \Delta^0  \bar{d}_j (g_{\bar{d}_j d_k \Delta^0} \gamma^5) d_k
.
\label{eq:ffS}
\end{align}
The couplings in the lepton sector are defined similarly to Eqs.~\eqref{eq:ffV} and \eqref{eq:ffS}.
The CP-even scalar couplings to the charged gauge bosons are 
\begin{align}
 {\cal L}_{VVh}
= &
  \sum_h g_{WWh} h W_{\mu}^{+} W^{- \mu} 
+ \sum_h g_{W'W' h} h W'^{+}_{\mu} W'^{- \mu}
+ \sum_h g_{WW' h} (h W_{\mu}^{+} W'^{- \mu} + h W'^{+}_{\mu} W^{- \mu})
,
\end{align}
where all the couplings are real.
The CP-odd scalar couplings to the gauge bosons are 
\begin{align}
{\cal L}_{VV\Delta} 
=&  
i \sum_{\Delta^0}g_{W W' \Delta^0}  W^{+}_{\mu} W'^{- \mu} \Delta^0 
-  i \sum_{\Delta^0}g_{W W' \Delta^0} W'^{+}_{\mu} W^{- \mu} \Delta^0 
.
\label{eq:VVDelta}
\end{align}
The structure of the triple gauge boson couplings are the same as in the SM case,
\begin{align}
{\cal L}_{gauge}
=&
\sum_{V^{\pm} = W, W} \sum_{V = \gamma, Z, Z'}
i g_{V^+ V^{-} V} 
\biggl(
(\partial_{\mu} V_{\nu}^{-} - \partial_{\nu} V_{\mu}^{-}) V^{+ \mu} V^{\nu} \nonumber\\
&+ (\partial_{\mu} V_{\nu}^{+} - \partial_{\nu} V_{\mu}^{+}) V^{\mu} V^{- \nu}
+ (\partial_{\mu} V_{\nu} - \partial_{\nu} V_{\mu}) V^{- \mu} V^{+ \nu}
\biggr).
\end{align}
The Feynman rules for the triple gauge bosons are obtained by replacing the coupling in the SM appropriately such as
$g_{WWZ}^{\text{SM}} \to g_{WWZ}$, $g_{WWZ}^{\text{SM}} \to g_{WW'Z}$.

Using the above interaction terms, 
we calculate the amplitude of $f_1 \bar{f}_2 \to V_3^{-} V_4^{+}$ in the high-energy limit.
We find that
\begin{align}
{\cal M}_{-+} 
=&
\frac{s}{2 m_{V_3} m_{V_4}} {\cal A} \sin\theta + {\cal O}(s^0)
,\\
{\cal M}_{+-} 
=&
\frac{s}{2 m_{V_3} m_{V_4}} {\cal B} \sin\theta + {\cal O}(s^0)
,\\
{\cal M}_{++} 
=&
\frac{\sqrt{s}}{2 m_{V_3} m_{V_4}} 
\left( {\cal C}^{(0)} + {\cal C}^{(1)} \cos\theta + {\cal O}(s^0) \right)
,\\
{\cal M}_{--} 
=&
\frac{\sqrt{s}}{2 m_{V_3} m_{V_4}} 
\left( {\cal D}^{(0)} + {\cal D}^{(1)} \cos\theta + {\cal O}(s^0) \right)
,
\end{align}
where $\theta$ is the angle between the two momenta of $f_1$ and $V_3^{-}$, 
$\sqrt{s}$ is the center-of-mass energy, and 
\begin{align}
{\cal A}
=&
- \frac{1}{2} \sum_F g_{\bar{F} f_1 V_3^{+}}^L (g_{\bar{F} f_2 V_4^{+}}^L)^{*}
+ \frac{1}{2} \sum_F (g_{\bar{f_1} F V_4^{+}}^L)^{*} g_{\bar{f}_2 F V_3^{+}}^L
- \sum_V g_{V_3^{+} V_4^{-} V} g_{\bar{f}_2 f_1 V}^L
,\\
{\cal B} =& - \sum_V g_{V_3^{+} V_4^{-} V} g_{\bar{f}_2 f_1 V}^R
,\\
{\cal C}^{(0)}
=&
-\frac{1}{2} \sum_F g_{\bar{F} f_1 V_3^{+}}^L (g_{\bar{F} f_2 V_4^{+}}^L)^{*} m_{f_1}
-\frac{1}{2} \sum_F (g_{\bar{f}_1 F V_4^{+}}^L)^{*} g_{\bar{f}_2 F V_3^{+}}^L m_{f_1}
\nonumber\\
&
-
\sum_{V}
\frac{m_{V_3^{+}}^2 - m_{V_4^{-}}^2 }{
m_V^2
}
g_{V_3^{+} V_4^{-} V}
( g_{\bar{f}_2 f_1 V}^L m_{f_1} - g_{\bar{f}_2 f_1 V}^R m_{f_2} )
\nonumber\\
& + \sum_{h} g_{V_3 V_4 h} g_{\bar{f}_2 f_1 h} 
\nonumber\\
&
+
 \sum_{\Delta^0}
 g_{\bar{f}_2 f_1 \Delta^{0}}
g_{V_3 V_4 \Delta^{0}}
\label{eq:C0}
,\\
{\cal D}^{(0)}
=&
+ \frac{1}{2} \sum_F g_{\bar{F} f_1 V_3^{+}}^L (g_{\bar{F} f_2 V_4^{+}}^L)^{*} m_{f_2} 
+ \frac{1}{2} \sum_F (g_{\bar{f}_1 F V_4^{+}}^L)^{*} g_{\bar{f}_2 F V_3^{+}}^L m_{f_2} 
\nonumber\\
&
+ 
\sum_{V}
\frac{m_{V_3^{+}}^2 - m_{V_4^{-}}^2 }{m_V^2}
g_{V_3^{+} V_4^{-} V}
( g_{\bar{f}_2 f_1 V}^R m_{f_1} - g_{\bar{f}_2 f_1 V}^L m_{f_2} )
\nonumber\\
&
- \sum_{h} g_{V_3 V_4 h} g_{\bar{f}_1 f_2 h} 
\nonumber\\
&
+  \sum_{\Delta^0}  g_{\bar{f}_1 f_2 \Delta^0} g_{V_3 V_4 \Delta^0} 
\label{eq:D0}
,\\
{\cal C}^{(1)} =& m_{f_1} {\cal A} + m_{f_2} {\cal B} ,\\
{\cal D}^{(1)} =& -m_{f_2} {\cal A} - m_{f_1} {\cal B}.
\end{align}
We obtain unitarity sum rules by imposing ${\cal A} = {\cal B} = {\cal C}^{(0)} = {\cal D}^{(0)} = 0$.
This condition is automatically satisfied in renormalizable models. 
Effective $V'$ models with a sufficiently high cutoff scale also satisfy this condition within a good approximation.
If ${\cal A} = {\cal B} = 0$, then ${\cal C}^{(1)}$ and ${\cal D}^{(1)}$ are automatically equal to zero. 
The sum rules from ${\cal C}^{(0)}$ and ${\cal D}^{(0)}$ contain both the CP-even and CP-odd couplings. 
We can separate these couplings by taking linear combinations, ${\cal C}^{(0)} \pm {\cal D}^{(0)}$. 
We finally obtain the following four independent unitarity sum rules.
\begin{align}
\sum_V g_{V_3^{+} V_4^{-} V} g_{\bar{f}_2 f_1 V}^L
=&
- \frac{1}{2} \sum_F g_{\bar{F} f_1 V_3^{+}}^L (g_{\bar{F} f_2 V_4^{+}}^L)^{*}
+ \frac{1}{2} \sum_F (g_{\bar{f_1} F V_4^{+}}^L)^{*} g_{\bar{f}_2 F V_3^{+}}^L
\label{eq:PU_2-1}
,\\
 \sum_V g_{V_3^{+} V_4^{-} V} g_{\bar{f}_2 f_1 V}^R =& 0
,\\
\sum_{\Delta^0}  g_{\bar{f}_2 f_1 \Delta^{0}} g_{V_3 V_4 \Delta^{0}}
=&
+ \frac{ m_{f_1} - m_{f_2}}{4} \sum_F 
\left(
g_{\bar{F} f_1 V_3^{+}}^L (g_{\bar{F} f_2 V_4^{+}}^L)^{*}  + (g_{\bar{f}_1 F V_4^{+}}^L)^{*} g_{\bar{f}_2 F V_3^{+}}^L
\right)
\nonumber\\
&
+
\frac{m_{f_1} + m_{f_2}}{2}
\sum_{V}
\frac{m_{V_3^{+}}^2 - m_{V_4^{-}}^2 }{m_V^2}
g_{V_3^{+} V_4^{-} V}
( g_{\bar{f}_2 f_1 V}^L  - g_{\bar{f}_2 f_1 V}^R )
\label{eq:PU_triplet}
,\\
\sum_{h} g_{V_3 V_4 h} g_{\bar{f}_2 f_1 h} 
=&
+ \frac{m_{f_1} + m_{f_2}}{4} \sum_F 
\left(
g_{\bar{F} f_1 V_3^{+}}^L (g_{\bar{F} f_2 V_4^{+}}^L)^{*} 
+
(g_{\bar{f}_1 F V_4^{+}}^L)^{*} g_{\bar{f}_2 F V_3^{+}}^L 
\right)
\nonumber\\
&
+
\frac{m_{f_1}- m_{f_2}}{2}
\sum_{V}
\frac{m_{V_3^{+}}^2 - m_{V_4^{-}}^2 }{m_V^2}
g_{V_3^{+} V_4^{-} V}
( g_{\bar{f}_2 f_1 V}^L  + g_{\bar{f}_2 f_1 V}^R)
.
\label{eq:PU_2-4}
\end{align}
In the following sections, we apply these sum rules to two simple setups
and discuss the relation between $\Gamma(W' \to WZ)$ and $\Gamma(W' \to ff)$.

\section{SM + $V'$ + CP-even scalars}
\label{sec:case1}
In this section, we apply the unitarity sum rules we found in the previous section to the following simple setup.
We consider SU(2)$_0 \times$SU(2)$_1 \times$U(1)$_2$ electroweak gauge symmetry.
Left-handed fermions are SU(2)$_1$ doublets.
Right-handed fermions are singlet under both SU(2)$_{0}$ and SU(2)$_{1}$.
Both the left- and right-handed fermions have appropriate U(1)$_2$ charge.
This charge assignment implies that the charged gauge bosons do not couple to the right-handed currents.
All scalars 
are CP even.
We do not include CP odd scalars in the setup here.
We do not specify the number of the CP-even scalars.
For simplicity, we assume the minimal flavor violation (MFV)~\cite{hep-ph/0207036}, 
namely all the flavor-changing structures are embedded in the CKM matrix.
This setup contains HVT model A~\cite{0907.5413,1402.4431}.
Thanks to these assumptions, the couplings in this setup are simplified as follows:
\begin{align}
&
 g_{\bar{u}_i d_j V^{+}}^{L} = V_{CKM}^{ij}  g_{V}, 
 \quad
 g_{\bar{\nu}_i \ell_j V^{+}}^{L} = \delta^{ij}  g_{V}^{\ell}, 
\\ 
&
 g_{\bar{u}_i u_j V^{0}}^{L,R} = \delta^{ij} g_{\bar{u} u V^{0}}^{L,R},
\quad
 g_{\bar{d}_i d_j V^{0}}^{L, R} = \delta^{ij} g_{\bar{d} d V^{0}}^{L, R},
\quad
 g_{\bar{\ell}_i \ell_j V^{0}}^{L, R} = \delta^{ij} g_{\bar{\ell} \ell V^{0}}^{L, R},
\\ 
&
 g_{\bar{u}_i u_j h}= \delta^{ij} g_{\bar{u} u h}, \quad
 g_{\bar{d}_i d_j h}= \delta^{ij} g_{\bar{d} d h}, \quad
 g_{\bar{\ell}_i \ell_j h}= \delta^{ij} g_{\bar{\ell} \ell h}
  .
\end{align}
Using these couplings, we can simplify the $u \bar{u} \to V^{-}_3 V^{+}_4$ unitarity sum rules 
as follows. 
\begin{align}
\frac{1}{2} g_{V_3} g_{V_4}
=&
\sum_{V = \gamma, Z, Z'} g_{V_3^{+} V_4^{-} V} g_{\bar{u} u V}^L
\label{eq:sum_L}
,\\
0 =&   \sum_{V = \gamma, Z, Z'} g_{V_3^{+} V_4^{-} V} g_{\bar{u}  u V}^R
\label{eq:sum_R}
,\\
g_{V_3} g_{V_4} 
=&  2 \sum_{h} g_{V_3 V_4 h}  \frac{g_{\bar{u} u  h}}{m_u}
\label{eq:sum_S1}
,\\
0 
=&
 \sum_{V = Z, Z'}
\frac{m_{V_3}^2 - m_{V_4}^2 }{m_V^2}
g_{V_3^{+} V_4^{-} V} ( g_{\bar{u} u V}^L  - g_{\bar{u} uV}^R ) 
\label{eq:sum_S3}
.
\end{align}
These unitarity sum rules are sufficient to discuss the ratio of $\Gamma(W' \to WZ)$ to $\Gamma(W' \to u_i d_j)$.
By combining Eqs.~(\ref{eq:sum_L}), (\ref{eq:sum_R}), and (\ref{eq:sum_S3}), 
and taking $V_3 = W$ and $V_4 = W'$, we find
\begin{align}
\frac{1}{2} g_W g_{W'} =& g_{WW'Z} (g_{\bar{u} u Z}^L - g_{\bar{u} u Z}^R) + g_{WW'Z'} (g_{\bar{u} u Z'}^L - g_{\bar{u} u Z'}^R)
\label{eq:srWW'1}, \\
g_{WW'Z'}  (g_{\bar{u} u Z'}^L - g_{\bar{u} u Z'}^R)  =& - \frac{m_{Z'}^2}{m_{Z}^2} g_{WW'Z} (g_{\bar{u} u Z}^L - g_{\bar{u} u Z}^R).
\label{eq:srWW'2}
\end{align}
Combining these unitarity sum rules, we can erase $g_{WW'Z'}$ and obtain
\begin{align}
g_{WW'Z} 
=  
-  \frac{m_{Z}^2}{m_{Z'}^2} \frac{g_W g_{W'} }{2 (g_{\bar{u} u Z}^L - g_{\bar{u} u Z}^R) } \frac{1}{1 - \frac{m_{Z}^2}{m_{Z'}^2} }.
\label{eq:coupling_ratio}
\end{align}
In general, $g_W$, $g_{\bar{u} u Z}^L$, and $g_{\bar{u} u Z}^R$ are different from the SM prediction 
but should become the same as in the SM in the decoupling limit ($m_{W', Z'} \to \infty$).
Thus, the relations among $g_W$, $g_{\bar{u} u Z}^L$, $g_{\bar{u} u Z}^R$, $m_W$, and $m_Z$ are the same as in the SM 
at the leading order in the large-$m_{W', Z'}$ limit. We find
\begin{align}
g_{WW'Z}  
\simeq
-  \frac{m_{W} m_{Z} }{m_{Z'}^2}  g_{W'}.
\label{eq:coupling_ratio_}
\end{align}
In a similar manner, we obtain the $\ell \bar{\ell} \to W^{-} W'^{+}$ perturbative unitarity sum rules and we find the following relation.
\begin{align}
g_{WW'Z}  
\simeq
-  \frac{m_{W} m_{Z} }{m_{Z'}^2}  g_{W'}^{\ell}.
\end{align}
By comparing this equation with Eq.~(\ref{eq:coupling_ratio_}), we find $g_{W'}^{\ell} \simeq g_{W'}$.
Thus, the relation given in Eq.~(\ref{eq:coupling_ratio_}) is flavor independent.
The partial widths for the $W'$ decays into $WZ$ and $ff$ are given by
\begin{align}
\Gamma(W' \to WZ) \simeq& \frac{1}{192 \pi} \frac{m_{W'}^5}{m_W^2 m_Z^2} g_{WW'Z}^2 , \\
\Gamma(W' \to u_i d_j ) \simeq& \frac{1}{16 \pi} |V_{CKM}^{ij}|^2m_{W'} g_{W'}^2, \\
\Gamma(W' \to \ell \nu ) \simeq& \frac{1}{48 \pi} m_{W'} g_{W'}^2,
\end{align}
where the terms of order $m_{W,Z,f}^2/m_{W'}^2$ have been neglected. 
From these equations, we find
\begin{align}
\frac{\Gamma(W' \to WZ)}{\Gamma(W' \to f_i f_j)}
 \simeq& 
 \frac{1}{4 c_{ij}}
 \frac{m_{W'}^4}{m_{Z'}^4} 
,
 \label{eq:complicated_R}
\end{align}
where 
\begin{align}
c_{ij} = 
\begin{cases}
 N_c |V_{CKM}^{ij}|^2 & \text{(for quarks)} \\
 \delta^{ij} & \text{(for leptons)}
\end{cases}
,
\end{align}
where $N_c =  3$.
Here we assume $g_{W'} \neq 0$. The case where $g_{W'} = 0$ is discussed in the end of this section.
The mass difference between $W'$ and $Z'$ depends on the amount of custodial SU(2) symmetry violation
due to the U(1)$_2$ gauge coupling and Yukawa couplings. 
We estimate that $| m_{Z'}^2  - m_{W'}^2| \simeq (g_2^2 \text{ and/or } y^2) v^2 \ll m_{W', Z'}^2$, where $v = 246$~GeV,
and thus 
\begin{align}
\frac{m_{W'}^2}{m_{Z'}^2} \simeq 1. \label{eq:mwp=mzp}
\end{align}
Therefore, Eq.~(\ref{eq:complicated_R}) is simplified as follows.
\begin{align}
\frac{\Gamma(W' \to WZ)}{\Gamma(W' \to f_i f_j)}
 \simeq& 
 \frac{1}{4 c_{ij}}.
 \label{eq:R}
\end{align}

The important consequence of Eq.~(\ref{eq:R}) is the suppression of Br$(W' \to WZ)$.
Since there are three generations in both the lepton and quark sectors,
\begin{align}
\frac{\Gamma(W' \to WZ)}{\sum_f \Gamma(W' \to f_i f_j)}
 \simeq& 
 \frac{1}{4 \times ( (N_c + 1) \times 3)}
= \frac{1}{48},
 \label{eq:R_sum}
\end{align}
where we use $\sum_{i, j} |V^{i j}_{CKM}|^2 = 3$.
This equation implies that Br($W' \to WZ$) $\lesssim$ 2\%. If we assume the equivalent theorem relation $\Gamma(W'\to WZ)\simeq \Gamma(W'\to Wh)$ and the $W'$ decay to $W$, and the heavy neutral scalar is highly suppressed, we find Br($W' \to WZ$) $\simeq$ 2\% and Br($W' \to e\nu$) $\simeq$ 8\% [$\sum_f$ Br($W' \to ff)$ $\simeq$ 96\%]. The assumption is justified in the case where $g_{WWh}\simeq g^{\rm{SM}}_{WWh}$ \cite{Abe:2015jra}. 
Therefore, the branching ratio of $W'$ to the gauge bosons is much smaller than its ratio to the fermions.
This result has been derived from the $f \bar{f} \to WW'$ unitarity sum rules
and does not need perturbative unitarity of other processes such as $WW \to WW$.
In addition, 
the result does not depend on the number of CP even scalars.
Therefore our result in this section can be applied to various models.

If Br($W' \to WZ)$ is measured in the future and if it is larger than 2\%, 
it is implied that the perturbative unitarity of $f \bar{f} \to WW'$ is violated
or
that other new particles in addition to $W', Z'$, and CP-even scalar bosons exist.

We briefly discuss the case where $g_{W'}=0$. 
In that case, $g_{WW'Z} = 0$, as we can see from Eq.~(\ref{eq:coupling_ratio}).
Therefore $W'$ decouples from the SM sector if $g_{W'} = 0$.

\section{SM + $V'$ + CP-even scalars + CP-odd scalars}
\label{sec:case2}
In this section, we extend the analysis in the previous section by introducing CP-odd scalars,
and show that Br($W' \to WZ$) can become larger than Br($W' \to ff$).
The CP-odd scalar couplings are given in Eqs.~(\ref{eq:ffS}) and (\ref{eq:VVDelta}).
As in the previous section, we assume MFV and CP conservation in the scalar sector.
Under these assumptions, all the CP-odd scalar couplings are 
simplified as follows:
\begin{align}
 g_{\bar{u}_i u_j \Delta^0} =&  +\frac{1}{2} g_{\bar{u} u \Delta^0 } \delta^{ij}, \quad
 g_{\bar{d}_i d_j \Delta^0} =   -\frac{1}{2} g_{\bar{d} d \Delta^0 } \delta^{ij}, \quad
 g_{\bar{\ell}_i \ell_j \Delta^0} =   -\frac{1}{2} g_{\bar{\ell} \ell \Delta^0 } \delta^{ij}.
\end{align}

We focus on the amplitude for $u \bar{u} \to W^{-} W'^{+}$ 
and obtain the same sum rules given in Eqs.~(\ref{eq:sum_L})--(\ref{eq:sum_S1}) again.
The unitarity sum rule in Eq.~(\ref{eq:sum_S3}) is modified by the contributions of the CP-odd scalars.
Instead of Eq.~(\ref{eq:sum_S3}), we obtain the following unitarity sum rule.
\begin{align}
\sum_{\Delta^0} \frac{g_{\bar{u} u \Delta^0}}{m_u}  g_{V_3 V_4 \Delta^0}
=&
 \sum_{V = Z, Z'}
2 g_{V_3 V_4 V}
\frac{m_{V_3}^2 - m_{V_4}^2 }{m_V^2}
 (g_{\bar{u} u Z}^L  - g_{\bar{u} u Z}^R )
.
\label{eq:ffVV_triplet}
\end{align}
The difference of this equation from Eq.~(\ref{eq:sum_S3}) is that
the left-hand side can be nonzero because of the contributions of the CP-odd scalars.
This is the only difference of the sum rules in this section from the previous section.
This difference can make $\Gamma(W' \to WZ)$ change drastically, as we will see in the following.
Using Eqs.~(\ref{eq:sum_L}), (\ref{eq:sum_R}), and (\ref{eq:ffVV_triplet}), we find that
\begin{align}
g_{WW'Z}
\simeq&
- \frac{m_W m_Z}{m_{Z'}^2} 
\left(
g_{W'} +  x_\Delta
\right)
,
\text{ where }
x_\Delta  = \sum_\Delta \frac{g_{\bar{u}u\Delta}}{m_u} \frac{g_{WW' \Delta}}{g_W}.
\label{eq:coupling_ratio_with_triplet}
\end{align}
Here we have neglected the terms of order $m_{W,Z,f}^2/m_{W'}^2$ 
and we have estimated that $m_{W'} \simeq m_{Z'}$, as we did in Eq.~(\ref{eq:mwp=mzp}).

We have two comments on Eq.~(\ref{eq:coupling_ratio_with_triplet}):
First, this equation is independent of the quark flavor,
although $x_\Delta$ looks quark-flavor dependent.
This is because $x_{\Delta} \simeq  -g_{W'}-g_{WW'Z}m^2_{Z'}/m_{W}m_{Z}$ and the right-hand side of this equation is
independent of quark flavor.
Second, Eq.~(\ref{eq:coupling_ratio_with_triplet}) implies that $g_{WW' Z} \neq 0 $ even if $g_{W'} = 0$ as long as the CP-odd couplings exist.
In the case where $g_{W'} = 0 $ and $g_{WW'Z} \neq 0$, 
the $W'$ decay to $WZ$ can be the dominant decay mode.
This is a big difference of the current setup from the setup discussed in Sec.~\ref{sec:case1}.

We find similar unitarity sum rules from the amplitude for $\ell \bar{\ell} \to W^{-} W'^{+}$.
The sum rule that corresponds to Eq.~(\ref{eq:coupling_ratio_with_triplet}) is given by
\begin{align}
g_{WW'Z}
\simeq&
- \frac{m_W m_Z}{m_{Z'}^2} 
\left(
g_{W'}^{\ell} +  x_\Delta^\ell 
\right),
\quad
\text{ where }
x_{\Delta}^\ell = \sum_\Delta \frac{g_{\bar{\ell} \ell \Delta}}{m_\ell} \frac{g_{WW' \Delta}}{g_W^{\ell}}.
\label{eq:coupling_ratio_with_triplet_lepton}
\end{align}
Unlike the setup discussed in Sec.~\ref{sec:case1}, we cannot conclude that $g_{W'}^{\ell} \simeq g_{W'}$ in this setup.

Using Eqs.~(\ref{eq:coupling_ratio_with_triplet}) and (\ref{eq:coupling_ratio_with_triplet_lepton}), we find
\begin{align}
\frac{\Gamma(W' \to WZ)}{\Gamma(W' \to u_i d_j)}
\simeq&
\frac{(g_{W'} +  x_\Delta)^2}{4 N_c |V_{CKM}^{ij}|^2 g_{W'}^2} 
\label{eq:R_with_Delta}
,\\
\frac{\Gamma(W' \to WZ)}{\Gamma(W' \to \ell \nu)}
\simeq&
\frac{(g_{W'}^{\ell} +  x_\Delta^{\ell})^2}{4 (g_{W'}^{\ell})^2} 
,\\
\frac{\Gamma(W' \to WZ)}{\sum \Gamma(W' \to f_i f_j)}
\simeq&
\left( 4 N_c \frac{3}{(1 +  \frac{x_\Delta}{g_{W'}})^2}   +   4 \frac{3}{(1 +  \frac{x_\Delta^{\ell}}{g_{W'}^{\ell}})^2} \right)^{-1}
\label{eq:R_with_Delta_sum}
,
\end{align}
where the terms of order $m_{W,Z,f}^2/m_{W'}^2$ have been neglected. 
The factor 3 in Eq.~(\ref{eq:R_with_Delta_sum}) is the number of the generation.
We use $\sum_{i, j} |V^{i j}_{CKM}|^2 = 3$.

We find that $g_{WW'Z}$ and $\Gamma(W' \to WZ)$ depend on $g_{W'}$, $x_\Delta$, and $x_\Delta^\ell$. 
This dependence is a different feature from Eqs.~(\ref{eq:R}) and (\ref{eq:R_sum}). 
If the CP-odd scalars are absent, 
then the ratio of the two partial decay widths is uniquely determined, as we have discussed in Sec.~\ref{sec:case1}; see Eq.~(\ref{eq:R_sum}). 
On the other hand, 
in the system with the CP-odd scalars, 
the ratio of the two partial decay widths is controlled by $g_{W'}$, $x_\Delta$, and $x_\Delta^{\ell}$, 
which are model-dependent parameters controlled by the CP-odd scalar couplings.
Thanks to this feature, $\Gamma(W' \to WZ)$ can be comparable to or even larger than the other decay modes.
For example, $\Gamma(W' \to WZ) \simeq \Gamma(W' \to \ell \nu)$ if $x_\Delta^\ell \simeq g_{W'}$ or $\simeq-3 g_{W'}$.
$\Gamma(W' \to WZ)$ is larger than $\Gamma(W' \to ff)$ in the large-$|x_\Delta/g_{W'}|$ regime.
$\Gamma(W' \to WZ)$ also can become small and even vanish for $x_\Delta \simeq x_\Delta^{\ell} \simeq -g_{W'}$.
In any case, the contributions of the CP-odd scalars significantly change the ratio 
of $\Gamma(W' \to WZ)$ to $\Gamma(W' \to ff)$ from the prediction without the CP-odd scalars.
In particular, it is an important feature that $W' \to ff$ is highly suppressed and $W' \to WZ$ can be the dominant decay mode in this setup with large $|x_\Delta/g_{W'}|$.
This is a very different feature from the setup in Sec. 3.

\begin{figure}[tb]
\includegraphics[width=0.9\hsize]{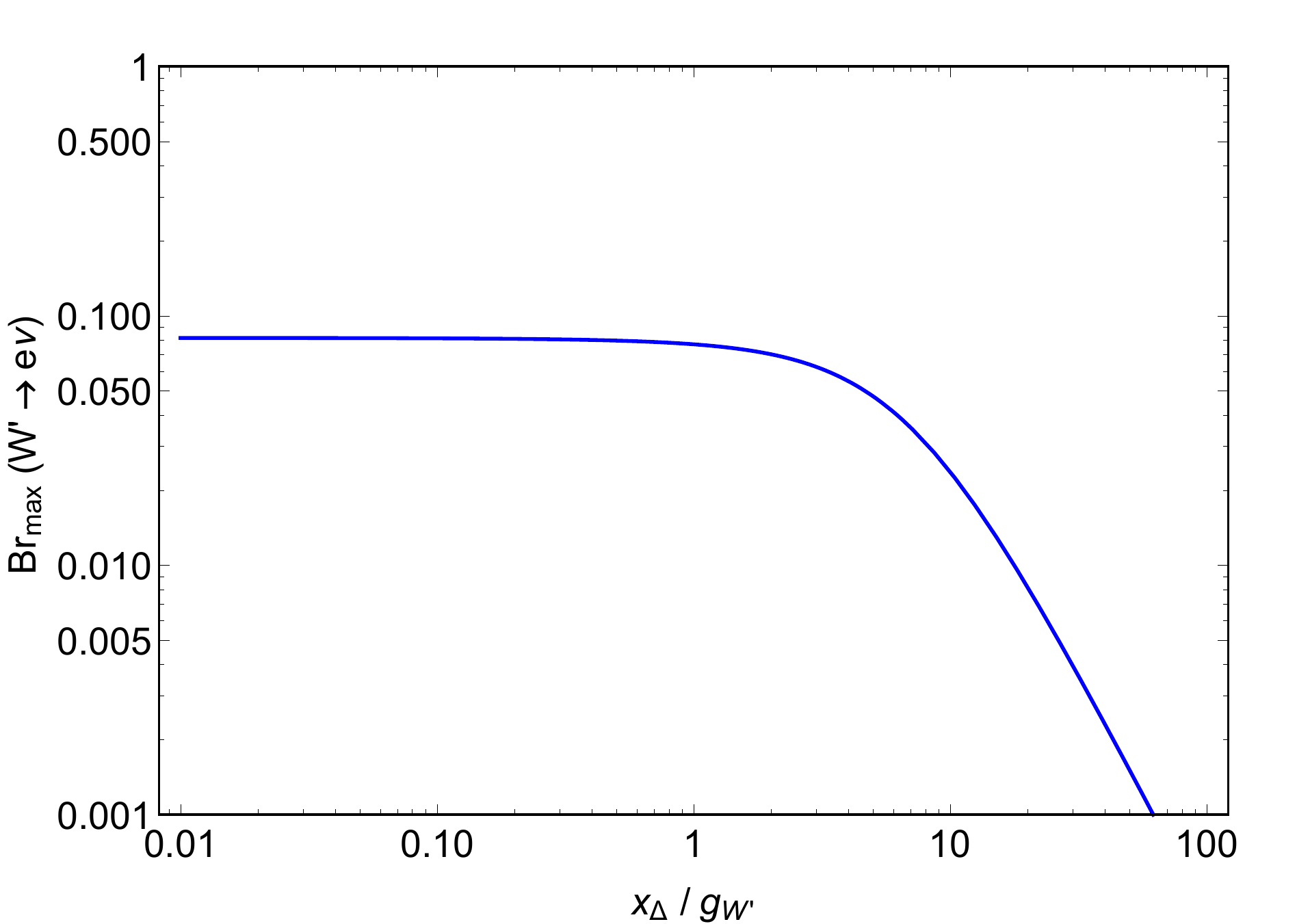}
\caption{
Br$_{\text{max}} (W' \to e\nu)$ as a function of $x_\Delta / g_{W'}$.
We can easily see the two extreme cases.
One is at the $x_{\Delta} \to 0$ limit where we can see Br$_{\text{max}} (W' \to e\nu)\simeq$ 8\%. 
This is consistent with the result in Sec.~\ref{sec:case1}.
The other case is at the $g_{W'}\to 0$ limit where Br$(W'\to ff)=0$ and $W'$ to $WZ$ can be the main decay mode.
This is again consistent with our discussion below Eq.~\eqref{eq:coupling_ratio_with_triplet}.
}
\label{fig:BrMax}
\end{figure}

We estimate the maximum value of Br($W' \to e\nu$) as follows:
\begin{align}
\text{Br}(W' \to e\nu)
=&
\cfrac{\Gamma(W' \to e\nu)}{\Gamma(W' \to WZ) + \sum_f \Gamma(W' \to ff) + \sum_X \Gamma(W' \to X)}
\nonumber \\
\leq &
\cfrac{\Gamma(W' \to e\nu)}{\Gamma(W' \to WZ) + \sum_f \Gamma(W' \to ff)}
\nonumber\\
\simeq &
\frac{4}{(1 + \frac{x_\Delta^\ell}{g_{W'}^\ell})^2}
\left( 1 + \frac{36}{(1 + \frac{x_\Delta}{g_{W'}})^2} + \frac{12}{(1 + \frac{x_\Delta^\ell}{g_{W'}^\ell})^2} \right)^{-1}
\nonumber\\
\equiv &
\text{Br}_{\text{max}}(W' \to e\nu)
,\label{eq:BrMax_enu}
\end{align}
where $\Gamma(W' \to X)$ is the sum of the other partial decay widths of $W'$
such as $\Gamma(W' \to Wh)$, $\Gamma(W' \to W \Delta^{0})$.
Br$_{\rm{max}}(W'\to e\nu)$ corresponds to the value of Br$(W'\to  e\nu)$ assuming $\Gamma(W'\to X)=0$. Although $\Gamma(W'\to X)$ generally takes nonzero value in the case where $m_{W'}>m_X$, 
Br$(W'\to  e\nu)$ is always smaller than Br$_{\rm{max}}(W'\to  e\nu)$ for any $x_\Delta/g_{W'}$, $x^l_\Delta/g_{W'}$, and $m_X$.
Figure~\ref{fig:BrMax} shows Br$_{\text{max}} (W' \to e\nu)$ as a function of $x_\Delta/g_{W'}$, assuming $x_\Delta^\ell = x_\Delta$
and $g_{W'}^\ell = g_{W'}$ for simplicity.
We find that Br$(W'\to e\nu)$ is highly suppressed and Br$(W'\to WZ)$ can be large in the case with large $x_\Delta/g_{W'}$. 
We can also see two extreme cases easily from this figure.
One is at the $x_{\Delta} \to 0$ limit where Br$_{\text{max}}(W'\to e\nu) \simeq$ 8\%. 
This result is consistent with the result in Sec.~\ref{sec:case1}.
The other case is at the $g_{W'}\to 0$ limit, where Br$(W'\to ff)=0$, and $W'$ to $WZ$ can be the main decay mode.
This result is again consistent with our discussion below Eq.~\eqref{eq:coupling_ratio_with_triplet}.
A negative $x_\Delta$ can also make Br($W' \to e\nu$) extremely small.
By measuring Br($W' \to e\nu$) and using 
Eqs.~(\ref{eq:coupling_ratio_with_triplet}), (\ref{eq:coupling_ratio_with_triplet_lepton}), and (\ref{eq:BrMax_enu}), 
we can estimate $x_\Delta$, and we can obtain information about the CP-odd scalar couplings even before the discovery of $\Delta^{0}$.

The existence of the CP-odd scalar couplings is important to the increase of Br$(W' \to WZ)$, 
because $x_\Delta$ and $x_\Delta^\ell$ can be zero if the CP-odd scalars do not couple to the fermions or to the gauge bosons,
as can be seen from Eqs.~(\ref{eq:coupling_ratio_with_triplet}) and (\ref{eq:coupling_ratio_with_triplet_lepton}).
Both $g_{\bar{f} f \Delta}$ and $g_{WW' \Delta^0}$ need to be nonzero 
in order to make Br($W' \to WZ)$ larger than 2\% by the effect of the CP-odd scalars.
To obtain a nonzero $g_{\bar{f} f \Delta}$, 
the scalar fields in the Yukawa terms need to contain CP-odd scalars.
For nonvanishing $g_{WW' \Delta^0}$,
the CP-odd scalars have to be components of scalar fields that develop vacuum expectation values (VEVs),
because $g_{WW' \Delta^0}$ originates from kinetic terms of the scalar fields.
These two conditions are useful guidelines to construct models that predict Br$(W' \to WZ) \gtrsim $ 2\%.

Let us discuss how to construct models that predict Br$(W' \to WZ) \gtrsim $ 2\%
by extending HVT model A.
It contains two scalar fields $H$ and $\Phi$, which are (\textbf{2, 1})$_{1/2}$
and (\textbf{2, 2})$_0$ under SU(2)$_0 \times $SU(2)$_1 \times $U(1)$_2$, respectively.
Since all the CP-odd scalars are eaten by the gauge bosons, we need to add other scalar fields to increase $\Gamma(W' \to WZ)$.
New scalar fields should not have the same gauge charge as $H$ and $\Phi$.
If there is more than one scalar field with the same gauge charge, 
we can redefine the scalar fields by taking their linear combination of them
and go to a basis where only one of the scalars develops a VEV.
This is equivalent to adding scalars that do not develop VEVs, and thus $g_{WW' \Delta^0} = 0$.
A simple choice for obtaining a nonzero $g_{WW' \Delta^0} $ is to add a scalar field that is (\textbf{1, 2})$_{1/2}$.
The model with this choice is discussed in Refs.~\cite{1305.2047,1507.01681}, 
and it certainly predicts a large Br$(W' \to WZ)$ with appropriate parameter choices.
Another possible choice is to add scalar fields that are representations of SU(2)$_{0}$ and/or SU(2)$_{1}$  larger than $\textbf{2}$,
for example, (\textbf{3, 3})$_0$.
However, such scalar fields break the custodial symmetry by their VEVs in many cases, 
and thus constraints should be studied carefully.

In this section, we have shown that
the CP-odd scalars can drastically change the ratio of $\Gamma(W' \to WZ)$ to $\Gamma(W' \to ff)$.
This result is the major difference from the result shown in Sec.~\ref{sec:case1}.
The difference originates from Eq.~\eqref{eq:PU_triplet}. 
The difference is very simple from the viewpoint of the unitarity sum rules,
although the phenomenological consequence drastically changes.

\section{Conclusions}
\label{sec:summary}

Spin-1 heavy vector bosons are popular particles predicted in models beyond the SM.
They decay into various particles, such as two SM gauge bosons.
In this paper, we have investigated the relation between two decay modes, $W' \to WZ$ and $W' \to ff$,
from the viewpoint of perturbative unitarity.
We have focused on the amplitudes of $f \bar{f} \to V^{-} V^{+}$, where $V = W, W'$,
and required that 
these processes do not have bad high-energy behavior at the tree level.
This requirement relates the couplings in the amplitudes to each other. 
The coupling relations obtained from this requirement are called the unitarity sum rules, which we have shown in Sec.~\ref{sec:PU}.
Using the unitarity sum rules, we can investigate 
the ratio of $\Gamma(W' \to WZ)$ to $\Gamma(W' \to ff)$.

In Sec.~\ref{sec:case1},
we have applied the unitarity sum rules to the system that contains
spin-1 heavy vector bosons and CP-even scalars as well as all the SM particles.
We have shown that $\Gamma(W' \to WZ)/\sum_f \Gamma(W' \to ff) \simeq 1/48$
where we sum over the contributions from three generations in both the quark and the lepton sectors.
Using this result, we have shown that Br$(W' \to WZ) \lesssim 2$\% in the system.
This result has been derived by imposing perturbative unitarity only on $f \bar{f} \to W^- W'^{+}$.
The same result is thus obtained even if perturbative unitarity is violated in other processes such as $WW \to WW$.
The result is independent of the number of CP-even scalars.
Moreover, the ratio of the two decay modes only depends on the color factor and $V_{CKM}$,
and thus is independent of details of models.
Hence we conclude that the result can be applied to various models.
If Br$(W' \to WZ)$ is measured in the future and is larger than 2\%, 
then perturbative unitarity requires new particles in addition to $V'$ and CP-even scalars.

In Sec.~\ref{sec:case2},
we have shown that CP-odd scalars help to increase Br$(W' \to WZ)$ 
if they couple to both the SM fermions and the SM gauge bosons.
In contrast to the case without the CP-odd scalars, 
Br$(W' \to WZ)$ depends on the parameters that are determined by the CP-odd scalar couplings
and can be much larger than 2\%. 
Depending on the couplings, 
the decay mode of $W'$ to the SM fermions is highly suppressed, and the decay mode of $W'$ to the gauge bosons can be dominant. 
This is a big difference of the models with CP-odd scalars from the models with only CP-even scalars. 
The measurement of the decay properties of $W'$ is thus important 
not only for understanding the property of $W'$ itself,
but also for revealing the structure of the system containing $W'$.
For example, we can estimate the strength of the CP-odd scalar couplings
to the SM particles from the measurement of Br$(W' \to e\nu)$ before the discovery of the CP-odd scalars. 
The result is also useful for model building.
For example,
we can see that the CP-odd scalars must be components of scalar fields that develop vacuum expectation values
in order to obtain large Br($W' \to WZ$), 
because
the nonzero CP-odd couplings to $W$ and $W'$ are required for large Br($W' \to WZ$)
and the couplings arise from the scalar kinetic terms.

The sum rule given in Eq.~\eqref{eq:PU_triplet} plays a crucial role in the prediction of Br($W' \to WZ$).
Only this sum rule contains the CP-odd scalar couplings
among the sum rules in Eqs.~\eqref{eq:PU_2-1}--\eqref{eq:PU_2-4}.
We have shown the importance of the CP-odd scalar couplings for Br$(W' \to WZ)$.
Br($W' \to WZ$) can be larger than 2\% if the left-hand side of Eq.~\eqref{eq:PU_triplet} does not vanish.
The sum rule given in Eq.~\eqref{eq:PU_triplet} is also important in setups
without the assumptions we made in this paper---namely, the minimal flavor violation and CP conservation in the scalar sector.
The analysis without these assumptions is straightforward but tedious, 
because it increases the number of parameters.
We leave this analysis for future work.

We have used the $f \bar{f} \to W^- W'^{+}$ unitarity sum rules in our analysis.
Our result does not change even if perturbative unitarity is violated in other processes such as $WW \to WW$.
Therefore, our result can be applied to various models.

\section*{Acknowledgments}
This work was supported by 
JSPS KAKENHI Grant No 16K17715 (T.A.)
and
Research Fellowships of the Japan Society for the Promotion of Science
(JSPS) for Young Scientists No. 263947 (R.N.).

\end{document}